\def\be{\begin{equation}}
\def\ee{\end{equation}}
\def\bea{\begin{eqnarray}}
\def\eea{\end{eqnarray}}

\documentclass[aps,groupedaddress,amsmath,amssymb,apsrev,superscriptaddress,showpacs,showkeys]{revtex4}
\usepackage{bm}
\usepackage{graphicx}
\begin{document}

\title{A study of $\mathcal{PT}$-symmetric Non-linear Schr\"odinger Equation}
\author{K.Nireekshan Reddy}
\affiliation{School of Physics, University Of Hyderabad,Hyderabad-500046,Andhra Pradesh,India}
\email{knireekshanreddy@gmail.com}
\author{Subhrajit Modak}
\affiliation{Indian Institute of Science Education and Reasearch (IISER) Kolkata, Mohanpur Campus, Mohanpur-741252}
\email{modoksuvrojit@iiserkol.ac.in}
\author{Kumar Abhinav}
\affiliation{Indian Institute of Science Education and Reasearch (IISER) Kolkata, Mohanpur Campus, Mohanpur-741252}
\email{kumarabhinav@iiserkol.ac.in}
\author{Prasanta K. Panigrahi}
\affiliation{Indian Institute of Science Education and Reasearch (IISER) Kolkata, Mohanpur Campus, Mohanpur-741252}
\email{pprasanta@iiserkol.ac.in}

\begin{abstract} 
Systems governed by the Non-linear Schr\"odinger Equation (NLSE) with various external $\mathcal{PT}$-symmetric
potentials are considered. Exact solutions have been obtained for the same through the method of ansatz, some 
of them being solitonic in nature. It is found that only the unbroken $\mathcal{PT}$-symmetric phase is realized
in these systems, characterized by real energies.
\end{abstract}

\pacs{03.65.Fd,11.30.Pb,11.30.Er}                           

\keywords{PT-symmetry,GP-equation.}

\maketitle

\section{Introduction}
In the recent past, following the initiation by Bender {\it et. al.} \cite{1,1a}, quantum systems with Hamiltonians
showing both parity ($\mathcal{P}$) and time-reversal ($\mathcal{T}$) or $\mathcal{PT}$-symmetry have been
attracting a lot of attention. They are characterized by real eigenvalues in certain parameter domain,
whereas complex-conjugate pairs of eigenvalues in other, separated respectively by whether the inherent
$\mathcal{PT}$-symmetry of the system is spontaneously maintained or broken. It immediately leads, both from
physical and mathematical perspectives, to the generalization of the usual Dirac-von Neumann scalar product
\cite{2}, in the same sense as that of the pseudo-Hermitian systems \cite{3}. These systems are characterized 
by a conserved non-local correlation, leading to unique boundary conditions \cite{4}, in agreement with
experimental results \cite{5}. Experimental realization of linear $\mathcal{PT}$-symmetric systems,
satisfying Schr\"odinger eigenvalue equation, have been made possible in optical analogues
\cite{Ru+10,Ko10,Gu+09,Re+12,Ma08,Mu+08a,Be08,Lo09,Ka+10,Ga07,We+10,LCV11,Be+10,KGM08}, where
$\mathcal{PT}$-symmetric dielectric constant substitute for potential under paraxial approximation.
On the other hand, non-linear complex systems have been studied, both analytically and numerically,
in various perspectives, {e.g.}, field theory \cite{6}, optical wave-guides \cite{Mu+08a}, lattice
dynamics \cite{8} and extensively for nonlinear Schr\"odinger equation (NLSE) with inherent
$\mathcal{PT}$-symmetry \cite{Mu+08a,9,10,11}. Interestingly for the $\mathcal{PT}$-symmetric NLSE
case, observed stable solutions, including solitons of bright and dark kind, are seen to represent
only the unbroken $\mathcal{PT}$-phase, with real energies. 
\paragraph*{} In this article, we analytically address a number of examples of imaginary/complex NLSE through the
method of ansatz, wherein interesting systems such as $\mathcal{PT}$-symmetric Scarf II potential leading to
solitonic solutions and harmonic $\mathcal{PT}$-symmetric potentials are considered. We restrict this discussion
to the cases, where only the linear or `trapping' part is $\mathcal{PT}$-symmetric, with rest of the NLSE having
real parameters, {\it e.g.}, coupling constant and chemical potential. It is to be mentioned here that, the
$\psi(\vec r,t)$ appearing in NLSE is {\it not} the usual Schr\"odinger eigenfunction, but rather the order parameter
obtained through ensemble averaging of a many-body wave-function of a second-quantized theory \cite{F}, and it also
contributes to the Hamiltonian $\hat H$ as $\vert\psi(\vec r,t)\vert^2$, thereby specifying $\hat H$ for each state. Thus,
the correspondence of $\mathcal{PT}$-symmetry of the eigenfunction and real/complex energies of linear quantum
mechanics is not intuitively carried over, though $\vert\psi(x,t)\vert^2$ still corresponds to density and there
is a well-defined energy of the system. Physically, along with dispersion and non-linearity of the usual NLSE,
the imaginary part of external/trapping potential provides dissipation, making a stable solution more difficult
to attain. Yet, it is observed that stable solutions with real energies are obtained, analogical to the
unbroken-$\mathcal{PT}$ phase of linear systems. However, no complex-conjugate energy spectrum, signifying broken
$\mathcal{PT}$-symmetry, exists in the present examples, subjected to the possible decay of such a system over
the ensemble-averaging over time. In other words, being real in nature, the non-linearity cannot balance the
imaginary contribution in the broken-$\mathcal{PT}$ phase. Thus, systems with self-coupling giving rise to
imaginary/complex coefficient of the nonlinear term in the NLSE  may retain a broken-$\mathcal{PT}$ phase,
which are out of context of the present article. Also, the stable solutions obtained here, with real energy and
$\mathcal{PT}$-symmetric $\psi(x,t)$, owe their existence to the face that dissipation due to imaginary part of the potential
is countered by the $\mathcal{PT}$-symmetry of the system itself, just as in $\mathcal{PT}$-symmetric Schr\"odinger
equation, which is a linear effect.
\paragraph*{} In the following, we introduce the concept of quasi one-dimensional (1-D) NLSE and its relevance 
with $\mathcal{PT}$-symmetry, which will be followed by one extensive example of the most general ansatz solution 
to the NLSE with $\mathcal{PT}$-symmetric Scarf II potential. Then, after listing a number of other examples and
successive discussion, we conclude.

\section{$\mathcal{PT}$-symmetric NLSE} 
The time-dependent NLSE in three dimension is given by:

\be
-\frac{\partial^2\Psi(\vec r,t)}{\partial x^2}+V(\vec r)\Psi(\vec r,t)+g\vert\Psi(\vec r,t)\vert^2\Psi(\vec r,t)-\mu\Psi(\vec r,t)=i\hbar\frac{\partial\Psi(\vec r,t)}{\partial t},~~~g,\mu\in\Re\label{NLSE1}
\ee
where $V(\vec r)$, $g$, $\mu$ and $E$ are trap potential, self-interaction coupling, chemical potential and
energy respectively, all being normalized as per $2m=1=\hbar$, $m$ being the effective mass parameter associated
with individual particles. In case of Bose-Einstein condensate (BEC), a
quasi-one-dimensional limit can be taken \cite{12} by factorizing as : $\Psi(\vec r)=\psi(x,t)G(y,z,\sigma)$,
$\sigma(x)$ being the local particle density along $x$-direction, assuming the trap to be harmonic in $y-z$
plane. Then, considering a well-defined energy to exist, we finally end up with the time-independent form:

\be
-\frac{\partial^2\psi(x)}{\partial x^2}+V(x)\psi(x)+g\vert\psi(x)\vert^2\psi(x)-\mu\psi(x)=E\psi(x).\label{NLSE2}
\ee
The above quasi 1-D NLSE is adopted for analyzing for the case when the external or trapping potential $V(x)$ 
is $\mathcal{PT}$-symmetric solely because the $\mathcal{P}$-operation is uniquely defined in 1-D only, and
most of the studies and understandings of $\mathcal{PT}$-symmetric systems have been obtained in 1-D. It also
is befitting as quasi 1-D NLSE systems are well-understood \cite{12,13}. We note that, the above system has
applicability to optical fibers \cite{GP}.
\paragraph*{} The $\mathcal{PT}$-symmetry is introduced in the above system by making $V(x)$ suitably complex,
{\t i.e.}. $V(x)=V_e(x)+iV_o(x)$, with $V_{e,o}\in\Re$ and the suffixes standing for `even' and `odd' respectively.
Below we discuss a number of such cases, starting with the extensive example of $\mathcal{PT}$-symmetric Scarf II
potential with the most generic ansatz, leading to solitonic solution under proper parameterization, followed by
some other interesting case. 
\subsection*{Example 1}
We consider the $\mathcal{PT}$-symmetric Scarf II potential, with general parametrization of the form:
\bea
V(x)&=&-A{\rm sech}^2(\alpha x)+i\alpha B{\rm sech}(\alpha x)\tanh(\alpha x),~~~A,B,\alpha\in\Re\nonumber\\
\text{with ansatz}:~~\psi(x)&=&a{\rm sech}(\alpha x)\exp{\left[ik\tan^{-1}\{\sinh(\alpha x)\}\right]},~~~a\in\Re~\text{or}~\Im,\label{E1}
\eea
yielding soliton. The corresponding terms in the NLSE are:
\bea
-\frac{\partial^2\psi(x)}{\partial x^2}&=&a\left[\left(\alpha^2+\alpha^2k^2\right){\rm sech}(\alpha x)-
\left(\alpha^2+\alpha^2+\alpha^2k^2\right){\rm sech}(\alpha x)\tanh^2(\alpha x)\right]\exp{\left[ik\tan^{-1}\{\sinh(\alpha x)\}\right]}\nonumber\\
&+&i\left(2 k\alpha^2+k\alpha^2\right) {\rm sech}^{2}(\alpha x)\tanh(\alpha x)\exp{\left[ik\tan^{-1}\{\sinh(\alpha x)\}\right]},\nonumber\\
V(x)\psi(x)&=&a\left[-A{\rm sech}^{3}(\alpha x)+i\alpha B{\rm sech}^{2}(\alpha x)\tanh(\alpha x)\right]\exp{\left[ik\tan^{-1}\{\sinh(\alpha x)\}\right]},~~~\text{\&}\nonumber\\
g\vert\psi(x)\vert^2\psi(x)&=&ga\vert a\vert^2{\rm sech}^{3}(\alpha x)\exp{\left[ik\tan^{-1}\{\sinh(\alpha x)\}\right]}
\eea
Then, on comparing the coefficients of linearly independent terms, $\tanh(\alpha x){\rm sech}^{2}(\alpha x)$, $\tanh^2(\alpha x){\rm sech}(\alpha x)$ and ${\rm sech}(\alpha x)$ respectively, the following `consistency conditions' are obtained:

\bea
i\left[3\alpha^2k+B\alpha\right]&=&0,\nonumber\\
(2+k^2)\alpha^2-A+g\vert a\vert^2&=&0,~~~\text{\&}\nonumber\\
(1+k^2)\alpha^2-\mu+g\vert a\vert^2-A&=&E.\label{C1}
\eea

\paragraph*{} The above conditions deliberately capture the ${\cal PT}$-symmetric phase of the exponent and thus the
same of the ‘wave-function’ $\psi(x)$. It is known that, for any power of the ${\rm sech}(\alpha x)$ term other than 1, 
solution of the NLSE with Scarf-II potential does not exist. Thus, only the ${\cal PT}$-symmetric sector is possible
for the above system. This is further confirmed from the last two equations of Eq. \ref{C1}, with corresponding energy
of the system as,

\be
E=-\alpha^2-\mu.\label{e1}
\ee
As the momentum $k$ is real, yielding $\mathcal{PT}$-symmetric wave-function with real energy. 
This is identifiable with the unbroken $\mathcal{PT}$-symmetric phase of linear systems. Only when $k$ is purely imaginary,
the wave-function is no longer $\mathcal{PT}$-symmetric, which is not possible for the above system.
\paragraph*{}Thus we conclude that, the broken-$\mathcal{PT}$ phase is not
allowed in this case, as it contradicts the initial conditions. This agrees with various theoretical and numerical
results in recent times \cite{Ma08,Mu+08a,11,14,15,16,17,18,19}, where only real-energy $\mathcal{PT}$-preserved
phase have been realized for systems with external complex potential only. The physical reason behind this is
suggested to be the instability of the mean-field ensemble-averaged solution against decay, signified by the
imaginary part of the energy in a non-linear system accounting for quantum fluctuations. However, as the coefficient
of the non-linear term, $g$, can be modified using Feshbach resonance \cite{20} and effectively through other
processes \cite{19}, bifurcations in the spectrum have been observed numerically, identifiable as the broken-$\mathcal{PT}$
phase. Here, we deal exclusively with systems having real $g$, which naturally arise in BEC and optical fibers,
and do not allow such phase.
\vskip 0.5cm
\paragraph*{{\bf The trigonometric counterpart:}} On considering the transformation $\alpha\rightarrow i\alpha$,
from the last example, we have the trigonometric (periodic) potential:

\bea
V(x)&=&-A\sec^2(\alpha x)+i\alpha B\sec(\alpha x)\tan(\alpha x),~~~A,B,\alpha\in\Re\nonumber\\
\text{with ansatz}:~~\psi(x)&=&a\sec(\alpha x)\exp{\left[ik\tan^{-1}\{\sin(\alpha x)\}\right]},~~~a\in\Re~\text{or}~\Im.\label{E2}
\eea
Proceeding as before, the consistency conditions turn out as:

\bea
i\left[3\alpha^2k-B\alpha\right]&=&0,\nonumber\\
(2-k^2)\alpha^2+A-g\vert a\vert^2&=&0,~~~\text{\&}\nonumber\\
-\alpha^2(1-k^2)-\mu+g\vert a\vert^2-A&=&E.\label{C2}
\eea
From the above, as $k$ is always real, again $\mathcal{PT}$-symmetry is maintained for the eigenfunction.
The $\mathcal{PT}$-preserved phase is further ensured by the real energy:

\be
E=\alpha^2-\mu.\label{e2}
\ee

\subsection*{Further Examples}
\vskip 0.5cm
{\bf I)}\hskip 0.25cm For the case:

\bea
V(x)&=&-A{\rm sech}^2(\alpha x)+i\alpha B\tanh(\alpha x),~~~A,B,\alpha\in\Re\nonumber\\
\text{with ansatz}:~~\psi(x)&=&a{\rm sech}(\alpha x)\exp{(ik\alpha x)},~~~a\in\Re~\text{or}~\Im,\label{E3}
\eea
the consistency conditions are:

\bea
i\left[2\alpha^2k+B\alpha\right]&=&0,\nonumber\\
-2\alpha^2+A-g\vert a\vert^2&=&0,~~~\text{\&}\nonumber\\
\alpha^2(1+k^2)-\mu-g\vert a\vert^2-A&=&E.\label{C3}
\eea
Here, the mandatory $\mathcal{PT}$-symmetric phase corresponds to the real energy:

\be
E=(k^2-1)\alpha^2-\mu,\label{e3}
\ee
with momentum $k$ being real.
\vskip 0.25cm
The trigonometric counterpart of the above system is:

\bea
V(x)&=&-A\sec^2(\alpha x)+i\alpha B\tan(\alpha x),~~~A,B,\alpha\in\Re\nonumber\\
\text{with ansatz}:~~\psi(x)&=&a\sec(\alpha x)\exp{(ik\alpha x)},~~~a\in\Re~\text{or}~\Im,\label{E4}
\eea
with consistency conditions:

\bea
i\left[2\alpha^2k-B\alpha\right]&=&0,\nonumber\\
-2\alpha^2-A+g\vert a\vert^2&=&0,~~~\text{\&}\nonumber\\
-(1-k^2)\alpha^2-\mu+g\vert a\vert^2-A&=&E.\label{C4}
\eea
The obtained phase is always $\mathcal{PT}$-symmetric, with energy:

\be
E=(k^2+1)\alpha^2-\mu,\label{e4}
\ee
with real $k$.

\vskip 0.5cm
{\bf II)}\hskip 0.25cm For the case:

\bea
V(x)&=&-A\sinh^{-2}(\alpha x)+i\alpha B\coth(\alpha x),~~~A,B,\alpha\in\Re\nonumber\\
\text{with ansatz}:~~\psi(x)&=&a\sinh(\alpha x)\exp{(ik\alpha x)},~~~a\in\Re~\text{or}~\Im,\label{E5}
\eea
the consistency conditions are:

\bea
i\left[2\alpha^2k+B\alpha\right]&=&0,\nonumber\\
-2\alpha^2+A+g\vert a\vert^2&=&0,~~~\text{\&}\nonumber\\
(1+k^2)\alpha^2-\mu-g\vert a\vert^2+A&=&E,\label{C5}
\eea
leading again to real momentum $k$, yielding real energy:

\be
E=(k^2-1)\alpha^2-\mu,\label{e5}
\ee
giving only the $\mathcal{PT}$-symmetric phase.

\vskip 0.25cm
The trigonometric counterpart of the above system:

\bea
V(x)&=&-A\csc^2(\alpha x)+i\alpha B\cot(\alpha x),~~~A,B,\alpha\in\Re\nonumber\\
\text{with ansatz}:~~\psi(x)&=&a\csc(\alpha x)\exp{(ik\alpha x)},~~~a\in\Re~\text{or}~\Im,\label{E6}
\eea
yields the consistency conditions:

\bea
i\left[2\alpha^2k+B\alpha\right]&=&0,\nonumber\\
-2\alpha^2-A+g\vert a\vert^2&=&0,~~~\text{\&}\nonumber\\
-(1-k^2)\alpha^2-\mu+g\vert a\vert^2-A&=&E,\label{C6}
\eea
leading to the $\mathcal{PT}$-symmetric phase of real energy:

\be
E=(k^2+1)\alpha^2-\mu.\label{e6}
\ee

\vskip 0.5cm
{\bf III)}\hskip 0.25cm Finally, we analyse a phase-locking form of ansatz solution for the system:

\bea
V(x)&=&-A{\rm sech}^2(x)+iB{\rm sech}(x)\tanh(x),~~~A,B\in\Re\nonumber\\
\text{with ansatz}:~~\psi(x)&=&ia{\rm sech}(x)+b\tanh(x),~~~a,b\in\Re,\label{E7}
\eea
leading to consistency conditions:

\bea
gab^2+Bb-Aa-2a-ga^3&=&0\nonumber\\
ga^2b+Ab-Ba+2b-gb^3&=&0\nonumber\\
1+a^2g+A-\mu&=&E\nonumber\\
gb^2-\mu&=&E,\label{C7}
\eea
which finally yields real energy:

\be
E=ga^2-\mu=gb^2-\mu,\label{e7}
\ee
for the $\mathcal{PT}$-symmetric phase.

\vskip 0.5cm
\section{Conclusion}
We have analyzed various examples of quasi 1-D NLSE with complex external potential having $\mathcal{PT}$-symmetry.
It was found that only the unbroken-$\mathcal{PT}$ phase is realized under most general parameterization, leading 
to real-energy solutions which are of both solitonic and periodic nature, and thus stable and experimentally
observable. A broken-$\mathcal{PT}$-phase may be realized through suitable complexification of the self-interaction
term, leading to the bifurcation in the complex energy space, observed in effective systems through numerical
study. This also opens-up a gateway to the interplay between quantum fluctuations captured by dissipation after
mean-field averaging.

\end{document}